\documentclass[aps,twocolumn,superscriptaddress,showpacs,prb,floatfix]{revtex4}
\usepackage{graphicx,rotating,subfigure,amsmath,amsfonts,amssymb,delarray}
\renewcommand{\vec}[1]{\boldsymbol #1}

\newcommand{\im}{\text{i}}
\def\l{\left}
\def\r{\right}
\def\12{\frac{1}{2}}
\def\nn{\nonumber}

\begin{document}
\bibliographystyle{apsrev}


\title{The $J_1-J_2$ Heisenberg model at and close to its $z=4$ quantum critical point}


\author{J. Sirker}
\affiliation{Department of Physics, Technical University Kaiserslautern, D-67663 Kaiserslautern, Germany}
\affiliation{Research Center OPTIMAS, Technical University Kaiserslautern, D-67663 Kaiserslautern, Germany}
\author{V. Y. Krivnov}
\affiliation{Joint Institute of Chemical Physics of RAS, Kosygin str.4,119334, Moscow, Russia}
\author{D. V. Dmitriev}
\affiliation{Joint Institute of Chemical Physics of RAS, Kosygin str.4,119334, Moscow, Russia}
\author{A. Herzog}
\affiliation{Department of Physics, Technical University Kaiserslautern, D-67663 Kaiserslautern, Germany}
\author{O. Janson}
\affiliation{Max Planck Institute for Chemical Physics of Solids, N\"othnitzer Str.~40, 01187 Dresden, Germany}
\author{S. Nishimoto}
\affiliation{IFW Dresden, P.O. Box 270116, D-01171 Dresden, Germany}
\author{S.-L. Drechsler}
\affiliation{IFW Dresden, P.O. Box 270116, D-01171 Dresden, Germany}
\author{J. Richter}
\affiliation{Institut fuer Theoretische Physik, Universitaet Magdeburg, P.O. Box 4120, D-39016, Magdeburg, Germany}


\date{\today}

\begin{abstract}
  We study the frustrated $J_1-J_2$ Heisenberg model with
  ferromagnetic nearest neighbor coupling $J_1<0$ and
  antiferromagnetic next-nearest neighbor coupling $J_2>0$ at and
  close to the $z=4$ quantum critical point (QCP) at $J_1/J_2=-4$. The
  $J_1-J_2$ model plays an important role for recently synthesized
  chain cuprates as well as in supersymmetric Yang Mills theories. We
  study the thermodynamic properties using field theory, a modified
  spin-wave theory as well as numerical density-matrix renormalization
  group calculations. Furthermore, we compare with results for the
  classical model obtained by analytical methods and Monte-Carlo
  simulations. 
  As one of our main results we present numerical evidence that the
  susceptibility at the QCP seems to diverge with temperature $T$ as
  $\chi\sim T^{-1.2}$ in the quantum case in contrast to the classical
  model where $\chi\sim T^{-4/3}$.
\end{abstract}

\maketitle
\section{Introduction}
\label{Intro}
\begin{figure}
\includegraphics*[width=0.9\columnwidth]{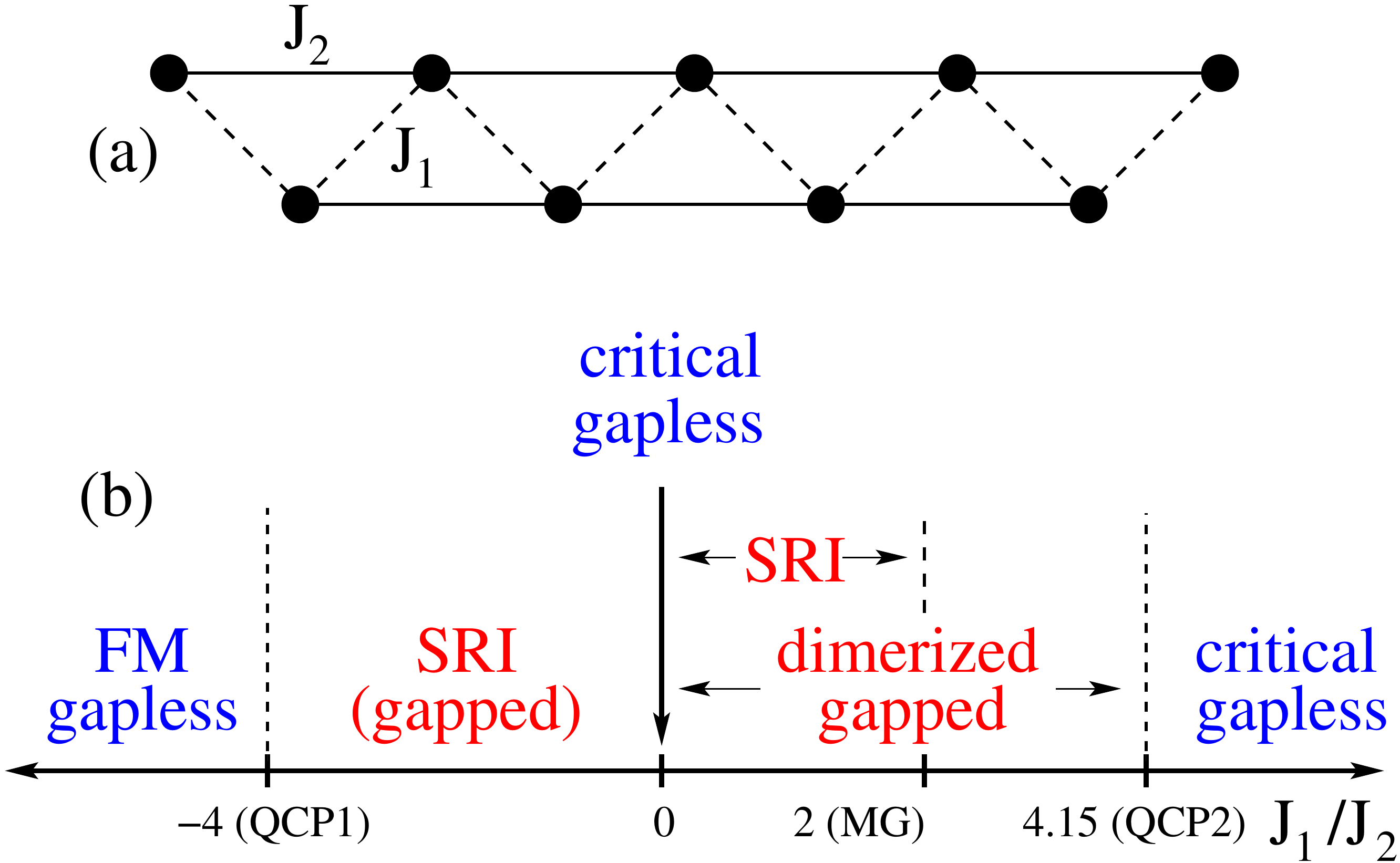}
\caption{(a) The $J_1-J_2$ chain viewed as a ladder with zigzag
  coupling. (b) Phase diagram of the $J_1-J_2$ chain as a function of
  $J_1/J_2$ with $J_2>0$.}
\label{J1J2.fig1}
\end{figure}

The frustrated one-dimensional (1D) $s=1/2$ Heisenberg model
\begin{equation}
\label{J1J2}
H=J_1\sum_j \vec{S}_j\vec{S}_{j+1} + J_2 \sum_j \vec{S}_j\vec{S}_{j+2}
\end{equation}
with nearest neighbor coupling $J_1$ and next-nearest neighbor
coupling $J_2$ is the minimal model to describe magnetism in a number
of cuprate chain compounds. It can be viewed as a
ladder with coupling $J_2$ along the legs and a zigzag rung coupling
$J_1$ as shown in Fig.~\ref{J1J2.fig1}(a).  Recently, there has been
renewed interest in this model with ferromagnetic coupling $J_1<0$ and
antiferromagnetic coupling $J_2>0$ propelled by the discovery of
multiferroic behavior in edge sharing spin
chains.\cite{DrechslerJMMM,MasudaZheludev,ParkChoi,SekiYamasaki,DrechslerVolkova,KatsuraNagaosa,Mostovoy}
The phase diagram of this model as a function of
$\tilde{\alpha}=J_1/J_2$ has been studied by a combination of field
theoretical and numerical
methods.\cite{TonegawaHarada2,ChitraPati,WhiteAffleck96,ItoiQin2,BursillGehring,NersesyanGogolin}
It is shown schematically in Fig.~\ref{J1J2.fig1}(b). At
$\tilde{\alpha}=0$ the system consists of two decoupled, critical
antiferromagnetic Heisenberg chains. By bosonization it has been found
that a small coupling, $|J_1|\ll 1$, leads to an exponentially small
gap
$\Delta\propto\exp(-\mbox{const}\,|\tilde{\alpha}|)$.\cite{WhiteAffleck96,ItoiQin2,NersesyanGogolin}
On the antiferromagnetic side, $\tilde{\alpha}>0$, the gapped phase
exists up to a critical point (QCP2 in Fig.~\ref{J1J2.fig1}(b)) at
$\tilde{\alpha}\approx 4.15$\cite{OkamotoNomura} 
where the system enters a critical gapless phase.
At the so-called
Majumdar-Ghosh (MG) point,\cite{MajumdarGhosh} $\tilde{\alpha}=2$, the
ground state is known exactly and consists of decoupled dimers.
Dimerization is indeed present for the whole gapped phase
$0<\tilde{\alpha}\lesssim 4.15$ while short-ranged incommensurate (SRI) spin
correlations have only been found for
$0<\tilde{\alpha}<2$.\cite{ChitraPati,WhiteAffleck96} On the ferromagnetic
side, $\tilde{\alpha}<0$, a phase with incommensurate spin-spin correlations
is followed by a ferromagnetic phase. The transition occurs at
$\tilde{\alpha}=-4$ (QCP1 in Fig.~\ref{J1J2.fig1}(b)), both in the quantum as
well as in the classical model.\cite{BaderSchilling,HaertelRichter}
Whereas the incommensurate (``spiral'') correlations are long-ranged
in the classical model, these correlations are predicted to be
short-ranged in the quantum model.\cite{NersesyanGogolin} However, the
gap is expected to be exponentially small and no numerical evidence
for this gap has been found yet.\cite{ItoiQin2,SirkerMF,SudanLuescher} 
At the critical point QCP1, the ferromagnetic state and states of
resonance valence bond (RVB) character are
degenerate.\cite{HamadaKane} In fact, all degenerate ground states at this
point can be explicitly constructed.\cite{SuzukiTakano1,SuzukiTakano2}
It turns out that there exists a unique ground state for a fixed total
spin $S_{\rm tot}$ and fixed $z$-component of the total spin, $S_{\rm
  tot}^z$. The ground state with $S_{\rm tot}^z=0$ is, in particular,
a uniformly distributed RVB state obtained as a superposition of all
possible states where sites are grouped in singlet pairs.

In this paper we will study 
the thermodynamic properties of 
the $J_1-J_2$ model near the quantum critical point QCP1. There are
two reasons why this model is of current interest: On the one hand,
the recently studied compound Li$_2$ZrCuO$_4$ has been shown to be
well described by the $J_1-J_2$ model with a frustration parameter
$\tilde{\alpha}$ putting the system into the spiral phase but rather
close to the critical point QCP1.\cite{DrechslerVolkova,SirkerMF} By
chemical or external presssure it might be possible to tune this or a
related system across the phase transition. On the other hand, it has
been shown using the anti de Sitter/conformal field theory (ADS/CFT)
correspondence that a deep connection between spin chains and string
theory exists.\cite{BeisertMinahan,BeisertKristjansen} In
$\mathcal{N}=4$ super Yang Mills theories the dilatation operator in
two-loop order can be represented as the $S=1/2$ spin chain,
Eq.~(\ref{J1J2}), with parameters fixed by the Yang Mills coupling
constant.\cite{Kruczenski,KruczenskiRyzhov} 
In the relevant parameter regime, both couplings $J_1$ and $J_2$ are
ferromagnetic in this case.  Interestingly, however, the second order
contribution taken separately has $J_1/J_2=-4$ although with $J_2<0$,
i.e., there is no frustration.  We will see in Sec.~\ref{NLSM} that
for this specific ratio---irrespective of the sign of $J_2$---certain
terms in the effective field theory will cancel exactly.

Our paper is organized as follows. In Sec.~\ref{NLSM} we present a
field theoretical description of the model in the ferromagnetic phase.
Based on this field theoretical model we will also discuss the
properties of the critical point QCP1 when approached from the
ferromagnetic side.
In Sec.~\ref{Classical} we investigate the thermodynamics of the
classical model. We test the analytical results for the
low-temperature properties obtained in Sec.~\ref{NLSM} by comparing
with Monte Carlo (MC) simulations. In Sec.~\ref{MSWT} analytical
results for the quantum model based on a modified spin-wave theory
(MSWT) are obtained.  We compare the MSWT predictions with the field
theory and with numerical data obtained by the density-matrix
renormalization group algorithm applied to transfer matrices (TMRG) in
Sec.~\ref{Numerical}. A summary and conclusions are presented in
Sec.~\ref{Concl}.

\section{Field theory and scaling arguments}
\label{NLSM}
We consider the case $\tilde{\alpha}<0$. Using spin coherent
states\cite{Fradkin} the Hamiltonian (\ref{J1J2}) can be mapped
onto a nonlinear sigma model with Euclidean action
\begin{eqnarray}
\label{NLSM.1}
S_E &=& -\im s\sum_r S_{WZ}[\vec{n}(r)] + S_H
\end{eqnarray}
with
(we set $\hbar=k_\text{B}=1$)
\begin{eqnarray}
\label{NLSM.2}
S_H &=& \int_0^\beta \!\!\! d\tau \sum_r \l\{J_1s^2\vec{n}(r,\tau)\vec{n}(r+a_0,\tau)\r. \nn \\
&+& \l. J_2s^2\vec{n}(r,\tau)\vec{n}(r+2a_0,\tau)\r\} \, .
\end{eqnarray}
Here $\vec{n}^2(r,\tau)=1$ is a unit vector, $s$ the spin quantum
number and $a_0$ the lattice constant. $S_{WZ}[\vec{n}(r)]$ is a
topological (Berry) term giving a phase which is determined
geometrically by the cap bounded by the trajectory $\vec{n}(\tau)$.
Without the topological term we have a classical action. Parametrizing
the unit vector in terms of angle variables and demanding that the
action is stationary we can easily find the classical ground state.
This leads to the well-known result that the ground state is
ferromagnetic for $\tilde{\alpha}<-4$ and a spiral with pitch angle
$\phi=\arccos(|\tilde{\alpha}|/4)$ for $\tilde{\alpha}>-4$.

Up to a constant we can replace $\vec{n}(r,\tau)\vec{n}(r+a_0,\tau)\to
-(\vec{n}(r,\tau)-\vec{n}(r+a_0,\tau))^2/2$ and similarly for the
next-nearest neighbor term. In the continuum limit we can then expand
the action in terms of the lattice constant $a_0$ and obtain in
leading orders
\begin{eqnarray}
\label{NLSM.3}
S_H &=& -\frac{J_2 s^2a_0}{2}\int_0^\beta \!\!\! d\tau \int_0^L dr (4+\tilde{\alpha})(\partial_r\vec{n})^2 \nn \\
&+&  \frac{J_2 s^2a_0^3}{24}\int_0^\beta \!\!\! d\tau \int_0^L dr (16+\tilde{\alpha})(\partial^2_r\vec{n})^2 \, ,
\end{eqnarray}
with $L=Na_0$ where $N$ is the number of lattice sites. 

It is instructive to briefly discuss the planar case where $\vec{n}$
is restricted to the $x$-$y$ plane. In this case we can parametrize
the unit vector by a single angle, $\vec{n}=(\cos\phi,\sin\phi,0)$,
leading to
\begin{eqnarray}
\label{NLSM.4}
S_H &=& -\frac{J_2 s^2a_0}{2}\int_0^\beta \!\!\! d\tau \int_0^L dr (4+\tilde{\alpha})(\partial_r\phi)^2  \\
&+&  \frac{J_2 s^2a_0^3}{24}\int_0^\beta \!\!\! d\tau \int_0^L dr (16+\tilde{\alpha}) \l[(\partial^2_r\phi)^2 +(\partial_r\phi)^4\r] \nn \, .
\end{eqnarray}
For $\tilde{\alpha}<-4$ we see that the action is minimized by
$\partial_r\phi = \partial^2_r\phi = 0$, i.e., the ground state is
ferromagnetic. For $\tilde{\alpha} > -4$ the system can gain energy by
forming a spin spiral.  Right at the transition point, the first line
of Eq.~(\ref{NLSM.4}) vanishes. The dispersion of the elementary
excitations (first term in the second line of Eq.~(\ref{NLSM.4}))
therefore becomes quartic at the critical point QCP1, $\omega_k\sim
k^4$. The critical theory therefore has a dynamical critical exponent
$z=4$ whereas in the ferromagnetic phase the dispersion is quadratic
($z=2$). The dynamical critical exponent relates the scaling of energy
$\omega$ and length $L$, $\omega\sim L^{-z}$. For the free energy
$f=-\frac{T}{L}\ln Z$ it follows that $f\sim -T^{3/2}$ in the
ferromagnetic phase and $f\sim -T^{5/4}$ at the critical point. The
same scaling relations apply for the inner energy. For the specific
heat $c=-T\partial^2f/\partial T^2$ it follows $c\sim T^{1/2}$ in the
ferromagnetic phase and $c\sim T^{1/4}$ at the critical point. These
scaling relations will stay valid also in the general case described
by Eq.~(\ref{NLSM.3}) and depend only on the dimension of the
dynamical critical exponent. 

Next, we consider the magnetic susceptibility in the ferromagnetic
phase. The operator in the second line of Eq.~(\ref{NLSM.3}) is then
irrelevant and can be ignored. The partition function, including a
magnetic field $h$, to leading order is then given by
\begin{eqnarray}
\label{NLSM.5}
Z &=& \int D\vec{n}\exp\l\{-\frac{1}{T}\int_0^L dr \l[\frac{\rho_s}{2}\l(\partial_r\vec{n}\r)^2 -hM_0 n^z\r]\r\} \nn \\
&=& \int D\vec{n}\exp\l\{-\int_0^{TL/\rho_s} dr' \l[\frac{\l(\partial_{r'}\vec{n}\r)^2}{2} -g n^z\r]\r\},
\end{eqnarray}
with the spin stiffness $\rho_s = -s^2a_0 (J_1+4J_2)$ and $M_0
=s/a_0$. In the second line we have rescaled $r'=Tr/\rho_s$ and
introduced a new parameter $g=hM_0\rho_s/T^2$. We are always
interested in $TL/\rho_s \gg 1$, i.e., in systems at temperatures $T$
much larger than the finite size gap $\sim 1/L$. If $g$ is the
only parameter of the theory, then we expect a universal scaling for
the magnetization, $M=M_0 \Phi(g)$, where $\Phi(g)\sim g
+\mathcal{O}(g^2)$ is a universal scaling function. For the
susceptibility it follows that $\chi\sim M_o^2\rho_s/T^2$.

Following Ref.~\onlinecite{TakahashiNakamura} one can even go one step
further and calculate the scaling function $\Phi(g)$ explicitly. To do
so it is important to realize that Eq.~(\ref{NLSM.5}) is nothing but the
imaginary time path integral of a quantum particle moving on a sphere.
The corresponding Hamiltonian is then given by $H=\vec{L}^2/2-gn^z$
where $\vec{L}$ is the angular momentum operator. 
The scaling function can now be obtained by calculating the
eigenspectrum of this Hamiltonian leading to
$\Phi(g)=\frac{2}{3}g+\mathcal{O}(g^3)$.\cite{TakahashiNakamura} If the scaling hypothesis is
valid, we expect the susceptibility at low temperatures on the
ferromagnetic side of the transition ($J_1+4J_2<0$) to be given by
\begin{equation}
\label{susci_ferro}
\chi = \frac{2}{3}\frac{M_0^2\rho_s}{T^2}=-\frac{2s^4}{3}\frac{J_1+4J_2}{T^2}=-\frac{2J_1s^4}{3T^2}\left(1+\frac{4}{\tilde{\alpha}}\right) \, .
\end{equation}
This relation has been found first in Ref.~\onlinecite{HaertelRichter}
based on an analysis of numerical data.  The low-temperature behavior
is therefore the same as for the nearest-neighbor ferromagnetic
Heisenberg model\cite{TakahashiNakamura} but with a rescaled spin
stiffness $\rho_s$. At the critical point QCP1 we have $\rho_s\to 0$
signalling the formation of spiral correlations. In our treatment we
have ignored the Berry phase term. In analogy to the simple
ferromagnetic model, we expect that this term 
in the ferromagnetic phase 
does not play any role for the low-temperature physics and thus
the low-temperature thermodynamic properties of the quantum and the
classical $J_1$-$J_2$ model are the same.

Let us now consider the field theory, Eq.~(\ref{NLSM.3}), at the
critical point
, still ignoring the topological term in Eq.~\ref{NLSM.1}.
The partition function is then given by
\begin{eqnarray}
\label{NLSM.6}
Z &=& \int D\vec{n}\exp\l\{-\frac{1}{T}\int_0^L dr \l[\tilde\rho_s\l(\partial_r^2\vec{n}\r)^2 -hM_0 n^z\r]\r\} \nn \\
&=& \int D\vec{n}\exp\l\{-\int_0^{L(T/\tilde\rho_s)^{1/3}}\!\!\!\!\!\!\! dr' \l[\l(\partial^2_{r'}\vec{n}\r)^2 -\tilde g n^z\r]\r\},
\end{eqnarray}
with $\tilde\rho_s = |J_1|s^2a^3_0/8$ and $\tilde g =
hM_0\tilde\rho_s^{1/3}/T^{4/3}$. The susceptibility therefore scales
as
\begin{equation}
\label{susci_crit}
\chi = C\frac{M_0^2\tilde\rho_s^{1/3}}{T^{4/3}}=C\frac{s^{8/3}}{2}\frac{|J_1|}{T^{4/3}} \, .
\end{equation}
where $C$ is a constant. Two of us have shown that one can again go
one step further by considering $Z$ as the path integral of a quantum
anharmonic oscillator.\cite{DmitrievKrivnov2010} The eigenvalues of
this Hamiltonian can then be calculated numerically. In full analogy
to the ferromagnetic case discussed before, the proportionality factor
can therefore be determined and is given numerically by $C\approx
2.14$.

The result (\ref{susci_crit}) is expected to be the exact low
temperature susceptibility for the classical model at $\tilde{\alpha}=-4$.
However, for the quantum model the theory is above the upper critical
dimension, $d+z=5$,
 with $d$ being the dimension of the system.
Spin-wave interaction terms might therefore yield
ultraviolet (UV) divergencies so that the result for the
susceptibility might not only depend on the parameter $\tilde g$ but also
on a UV cutoff.\cite{Sachdev} In this case the scaling hypothesis
would be violated and the formula (\ref{susci_crit}) not applicable
for the quantum model. 
Furthermore, the topological term which we have neglected throughout,
is likely to play an important role at QCP1. Here the ferromagnetic
state is degenerate with RVB states\cite{HamadaKane} which do not
exist for the classical model. Within the nonlinear sigma model
description one might expect that part of this difference is encoded
in a nontrivial topological term.
In the following, we will first check the analytical predictions for
the classical model before analyzing a possible violation of
Eq.~(\ref{susci_crit}) for the quantum model.

\section{The classical model}
\label{Classical}
The classical nearest-neighbor ferromagnetic Heisenberg model has been
solved by Fisher [\onlinecite{Fisher}]. For the classical model
(\ref{J1J2}) with antiferromagnetic $J_1$ and $J_2$, Harada and
Mikeska [\onlinecite{HaradaMikeska}] have shown that thermodynamic
quantities can be expressed in terms of eigenvalues of transfer
matrices which follow from integral equations. 
The scaling $\chi\sim T^{-4/3}$ at the Lifshitz point,
$\tilde\alpha=-4$, has already been discussed in
Ref.~[\onlinecite{Selke}].
Recently, also the thermodynamics for general $J_1<0$ and $J_2>0$ has
been studied in more
detail.\cite{DmitrievKrivnov1,DmitrievKrivnov2,DmitrievKrivnovRichter,DmitrievKrivnov2010,DmitrievKrivnov3}

In the classical case, the results derived in the previous section by
field theory methods---both for the ferromagnetic phase and the
critical point---should be valid. Here we concentrate on providing
numerical evidence that the parameter-free formulas for the magnetic
susceptibility, Eq.~(\ref{susci_ferro}) and (\ref{susci_crit}),
respectively, are correct. The numerical data are obtained using
Monte Carlo simulations from the ALPS package\cite{ALPS} with a
cluster update and a system size $N=10000$.

\begin{figure}
\includegraphics*[width=1.0\columnwidth]{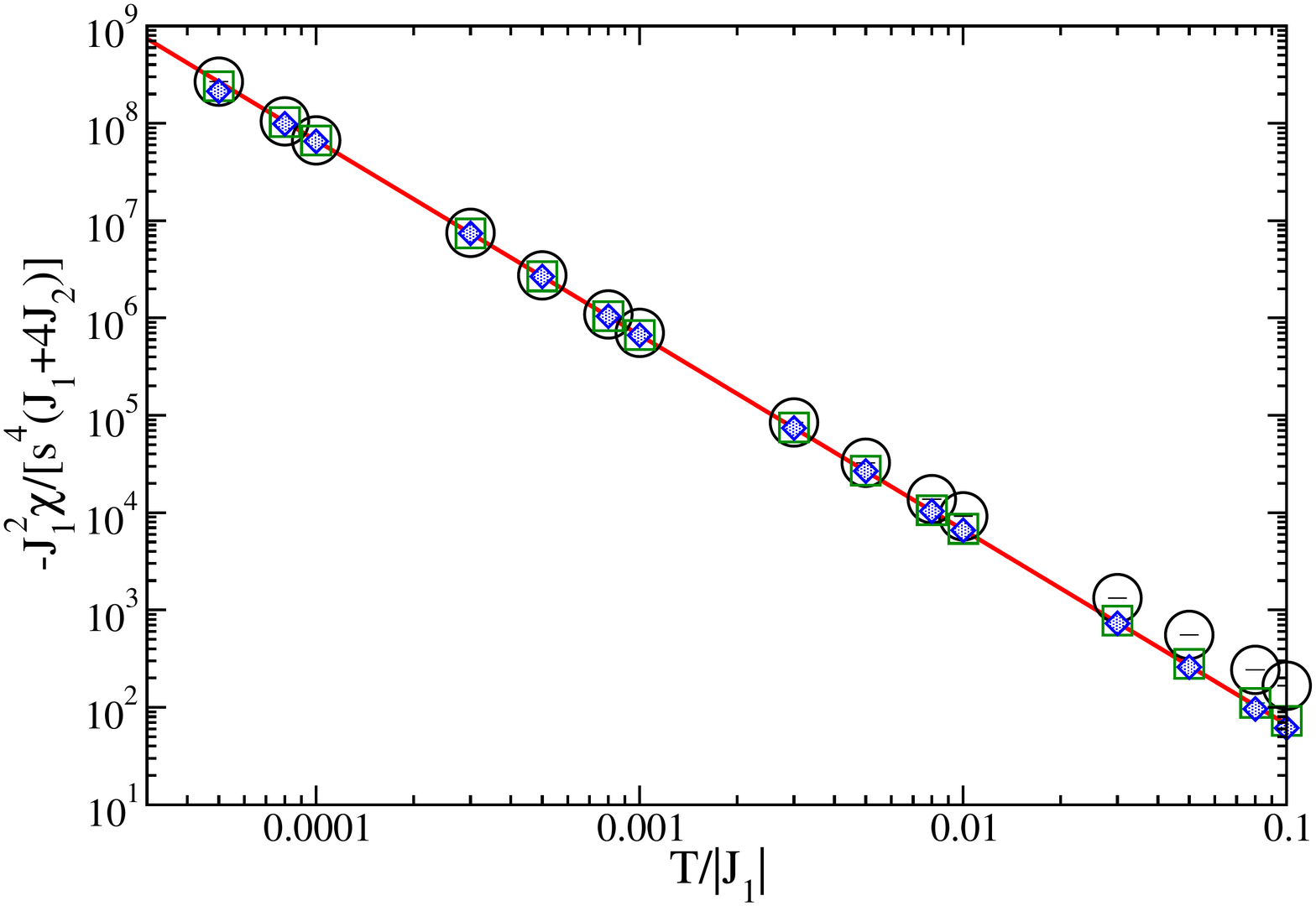}
\caption{(Color online) Universal data collapse,
  $\chi/[s^4(-J_1-4J_2)]=2/(3T^2)$, at low temperatures for the
  classical model. The line denotes the analytical result
  (\ref{susci_ferro}), and the symbols MC results for
  $\tilde{\alpha}=-5$ (circles), $\tilde{\alpha}=-10$ (squares), and
  $\tilde{\alpha}=-20$ (diamonds).}
\label{MC_ferro}
\end{figure}

For the ferromagnetic phase, Eq.~(\ref{susci_ferro}) predicts that at
low temperatures all data for $\chi/[s^4(J_1+4J_2)]$ should collapse
onto a single universal curve. In Fig.~\ref{MC_ferro} Monte Carlo
results for various $\tilde{\alpha}$ are compared with the analytical
formula.
The data collapse onto the analytical curve is perfect over
temperatures of several orders of magnitude. We note that the closer
$\tilde{\alpha}$ is to the critical point QCP1 the lower the
temperatures are where the universal scaling sets in.

\begin{figure}
\includegraphics*[width=1.0\columnwidth]{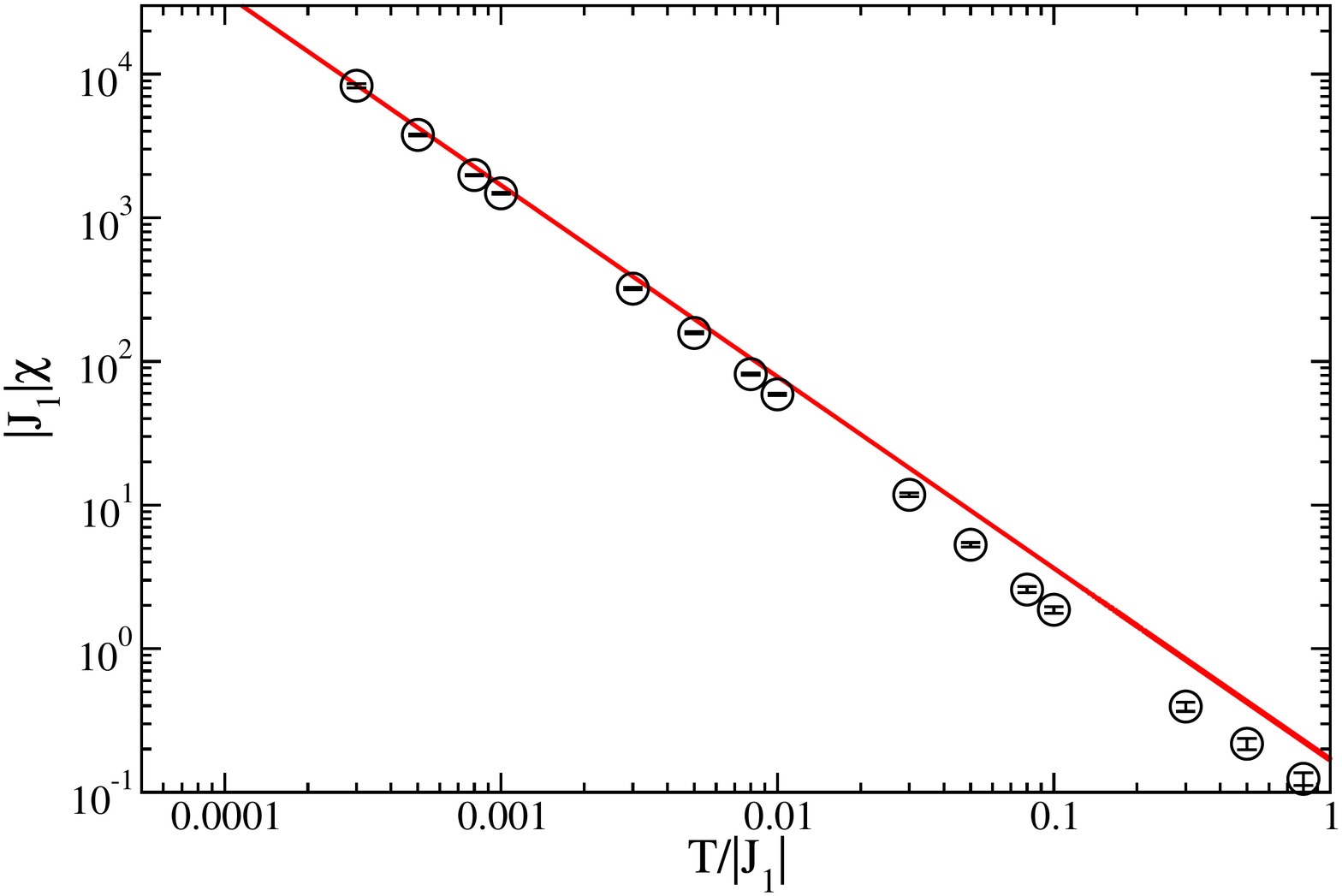}
\caption{(Color online) Analytical formula (\ref{susci_crit}) for the
  low-temperature susceptibility at $\tilde{\alpha}=-4$ (line) compared to MC
  data (circles).}
\label{MC_QCP}
\end{figure}

Similarly, we can also check the formula for the critical point,
$\tilde{\alpha}=-4$, see Fig.~\ref{MC_QCP}.
The numerical data do confirm the analytical result, however, we note
that temperatures $T/|J_1|\ll 0.01$ are required.

\section{The quantum model}
The quantum $s=1/2$ nearest-neighbor ($J_2=0$) ferromagnetic
Heisenberg chain is exactly solvable. Thermodynamic properties, in
particular the susceptibility, have been calculated using the
thermodynamic Bethe ansatz.\cite{TakahashiMSWT4,TakahashiMSWT} The
results have been shown to be in excellent agreement with those
obtained by a modified spin-wave theory (MSWT).\cite{TakahashiMSWT} In
the following we extend the MSWT approach to the ferromagnetic phase
of the $J_1$-$J_2$ model and to the quantum critical point. We then
test the obtained analytical results by comparing with numerical data
obtained by the transfer matrix renormalization
group.\cite{BursillXiang,WangXiang,SirkerKluemperPRB,SirkerKluemperEPL,SirkerMF}

\subsection{Modified spin wave theory}
\label{MSWT}
To calculate the thermodynamic properties of the spin-$s$
ferromagnetic Heisenberg chain, Takahashi
[\onlinecite{TakahashiMSWT,TakahashiMSWT4}] introduced a modified
spin-wave theory. The spin operators are represented by bosonic
operators as in regular spin-wave theory. In addition, the constraint
of vanishing magnetization at finite temperatures posed by the
Mermin-Wagner theorem is implemented in a simple way by adding an
effective magnetic field which acts as a Lagrange multiplier for the
magnetization. MSWT has also been used successfully to describe
boundary contributions in the open ferromagnetic Heisenberg chain
\cite{SirkerBortz} as well as the thermodynamics in the dimerized
ferromagnet.\cite{SirkerHerzog, HerzogSirker, SirkerHerzog2} Moreover
it has been shown that the classical ferromagnetic chain is
well-described by MSWT.\cite{TakahashiMSWT,SirkerBortz,Takahashi2DFM}
Here we will apply the same method to the Hamiltonian (\ref{J1J2})
with general spin $s$.

We expect that MSWT can be applied for $\tilde{\alpha}< -4$ where the ground
state is ferromagnetic. We will also use this approximation for the
QCP at $\tilde{\alpha}=-4$, however, here the validity of MSWT is
questionable
because of the degeneracy of the ferromagnetic ground state with RVB
states.
A further discussion will be presented in Sec.~\ref{Numerical}, based
on a comparison with numerical data, and in the conclusions,
Sec.~\ref{Concl}. In the MSWT approximation the Hamiltonian
(\ref{J1J2}) is represented as
(in the following we set the lattice constant $a_0=1$)
\begin{equation}
\label{MSWT.1}
	H=N(J_1+J_2)s^2+\sum_k \omega_k a^\dagger_k a_k
\end{equation}
where $a_k^{(\dagger)}$ is a bosonic annihilation (creation) operator
with $[a_k,a_{k'}^\dagger]=\delta_{k,k'}$. The dispersion relation is
given by $\omega_k=2s(|J_1|\left[1-\cos
  k\right]+J_2\left[\cos(2k)-1\right])$. The additional constraint of
vanishing magnetization reads $s=N^{-1}\sum_k n_k$ where
$n_k=(\exp[\omega_k/T+v]-1)^{-1}$ is the Bose function
including the effective magnetic field, $v\equiv h/T$. 

First, we study the case $\tilde{\alpha}<-4$. For small temperatures only
spin-wave excitations with small momenta contribute and the dispersion
can be approximated as $\omega_k=|J_1|s(1-4/|\tilde{\alpha}|)k^2$.  The
constraint can now be solved explicitly by expanding in the reduced
temperature $t$. This leads to
\begin{eqnarray}
\label{MSWT.2}
	\sqrt{v}&=&\frac{\sqrt{t}}{2s}\left(1+\frac{\zeta(\frac{1}{2})}{\sqrt{\pi}}\frac{\sqrt{t}}{2s}+\frac{\zeta^2(\frac{1}{2})}{\pi}\l(\frac{\sqrt{t}}{2s}\r)^2+\dots\right), \nn \\
t&\equiv&\frac{T}{|J_1|s(1-4/|\tilde{\alpha}|)} .
\end{eqnarray}
Here $\zeta(x)$ is the Riemann zeta-function. We note that the
expression for $v$ is the same as for the simple
ferromagnet,\cite{TakahashiMSWT} only the definition of the reduced
temperature is modified. The expansion (\ref{MSWT.2}) is valid if
$\sqrt{t}/2s\ll 1$, i.e., the temperature range where this result is
applicable shrinks the closer we get to the critical point. The free
energy in spin-wave approximation is given by
\begin{equation}
\label{MSWT.3}
f=(J_1+J_2)s^2-T\l[vs+\frac{1}{N}\sum_k\ln(1+n_k)\r] \, .
\end{equation}
At small temperatures we can expand the second term in $v$ and obtain
\begin{equation}
\label{MSWT.4}
f=(J_1+J_2)s^2-T\l[\frac{\zeta(3/2)}{2}\sqrt{\frac{t}{\pi}}-\frac{t}{4s}+\cdots\r] \, .
\end{equation}
In agreement with the scaling relations derived in Sec.~\ref{NLSM} we
find as leading temperature dependence $f\sim -T^{3/2}$. From
$C=-T\partial^2f/\partial T^2$ the leading temperature dependence of
the specific heat can be obtained.

The susceptibility in MSWT is given by
\begin{equation}
\label{MSWT.5}
\chi=\frac{1}{3TN}\sum_k n_k(n_k+1)  
\end{equation}
leading to the low-temperature expansion
\begin{equation}
\label{MSWT.6}
\chi=\frac{s^3}{T}\left[\frac{2}{3t}-\frac{\zeta(\frac{1}{2})}{\sqrt{\pi s^2t}}+\dots\right] \, .
\end{equation}
The leading temperature dependence found in MSWT therefore agrees
exactly with formula (\ref{susci_ferro}) found by general scaling
arguments.

Next, we consider the critical point $\tilde{\alpha}=-4$. For small momenta
the dispersion now reads $\omega_k=|J_1|sk^4/4$ and thus  the dispersion
changes from quadratic to quartic which will have consequences for the
temperature scaling of thermodynamic quantities. For the constraint we
find
\begin{equation}
\label{MSWT.7}
v=\l(\frac{\tilde t}{64s^4}\r)^{1/3}\l[1+\frac{\sqrt{2}\zeta(1/4)}{3s\Gamma(3/4)}\tilde t^{1/4}+\cdots\r] 
\end{equation}
with a new reduced temperature $\tilde t\equiv 4T/(|J_1|s)$. Using
relation (\ref{MSWT.3}) we obtain for the free energy
\begin{equation}
\label{MSWT.8}
f=(J_1+J_2)s^2-T\l[\frac{\Gamma(1/4)\zeta(5/4)}{4\pi}\tilde t^{1/4}-3\l(\frac{\tilde t}{64s}\r)^{1/3}\r] 
\end{equation}
plus higher order terms. The scaling of the leading term, $f\sim
-T^{5/4}$, is again consistent with the scaling arguments in
Sec.~\ref{NLSM}.  Finally, we can calculate the susceptibility using
Eq.~(\ref{MSWT.5}) and obtain
\begin{equation}
\label{MSWT.9}
	\begin{aligned}
		\chi=\frac{1}{T}\left[\left(\frac{s^7}{\tilde t}\right)^{1/3}-\frac{7\zeta(\frac{1}{4})s^{4/3}}{6\sqrt{2}\Gamma(\frac{3}{4})}\left(\frac{1}{\tilde t}\right)^{1/12}+\frac{S}{3}+\dots\right]
	\end{aligned}
\end{equation}
We note that the scaling of the leading term is the same as in
Eq.~(\ref{susci_crit}). However, the numerical prefactor is not the
same.  For the case $s=1/2$ we find, in particular, from
(\ref{MSWT.9}) that $\chi=2^{-10/3}T^{-4/3}\approx 0.0992\,
T^{-4/3}$.
\footnote{In Ref.~[\onlinecite{DmitrievKrivnov2010}] the result
  $\chi=(3/16)T^{-4/3}$ was obtained within MSWT by a direct
  minimization of the free energy functional. However, this approach
  gives rise to a term $\sim k^2$ in the dispersion which seems to be
  unphysical. We therefore believe that the treatment presented here,
  where the dispersion remains quartic, is more accurate.} 
In contrast, Eq.~(\ref{susci_crit})---which does give the correct
low-temperature behavior of the classical model, see
Fig.~\ref{MC_QCP}---yields $\chi\approx 0.1685\, T^{-4/3}$. First of
all, this does suggest that at the QCP the classical and the quantum
model no longer show the same thermodynamic properties at low
temperatures. However, similarly to the scaling approach used in
Sec.~\ref{NLSM} one might also question the foundations of MSWT for
the QCP alltogether. At the QCP the ground state is no longer a simple
ferromagnet suggesting that also the excitations are no longer
described by simple spin waves. Furthermore, UV divergencies in
contributions from spin wave interaction terms might be expected
because we are now dealing with a theory above the upper critical
dimension. An independent test of the scaling and the MSWT approach at
the QCP can only be obtained by unbiased numerical calculations. Such
calculations will be presented in the next subsection.

\subsection{Numerical results}
\label{Numerical}
\begin{figure}
\includegraphics*[width=.95\columnwidth]{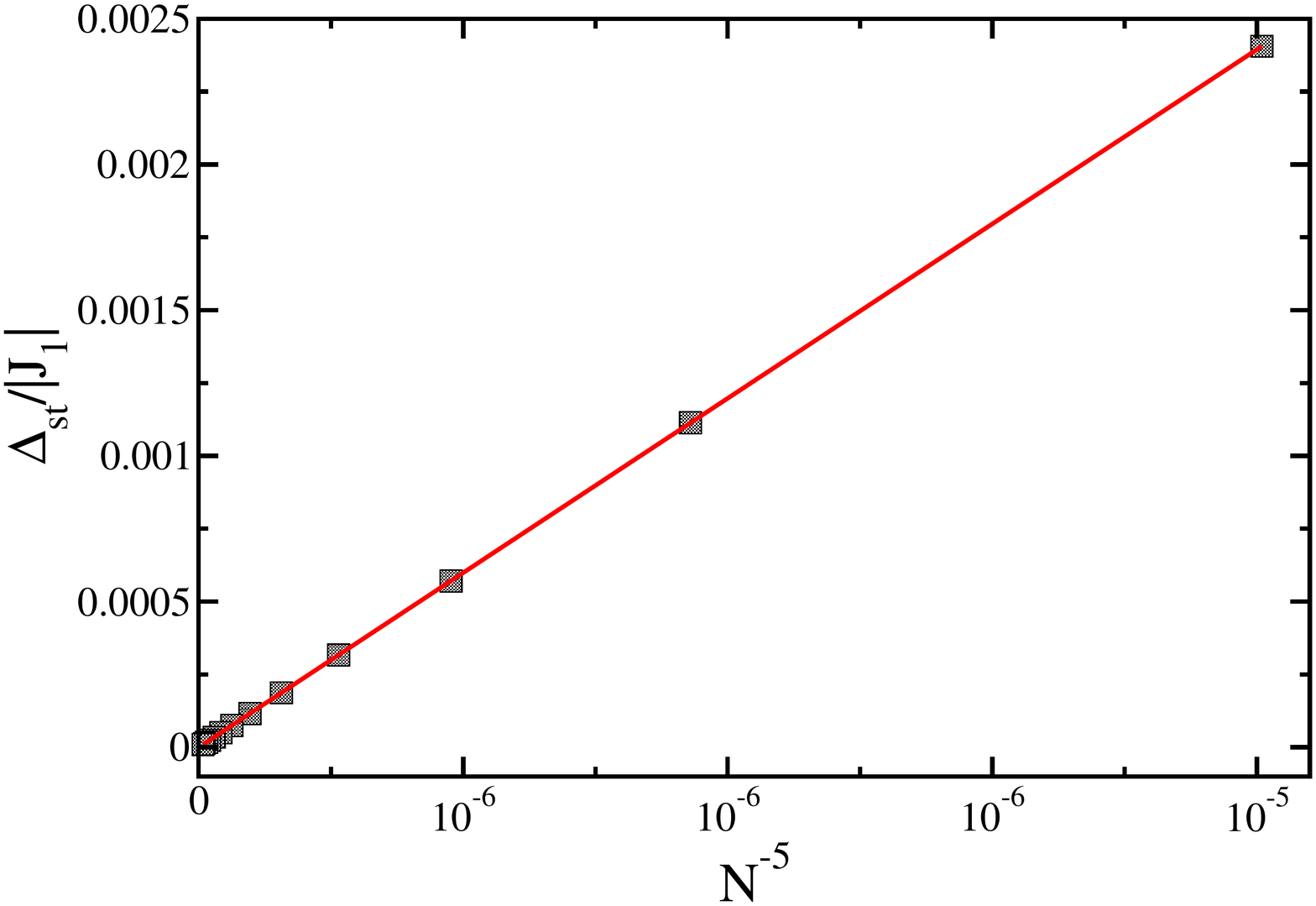}
\caption{(Color online) Numerical data (squares) for the
  singlet-triplet gap $\Delta_{st}$ for chains of even lengths
  $N=12,14,\cdots,36$. The solid line is a fit
  $\Delta_{st}=a_0+a_1\l(N^{-5}\r)^{a_2}$ with $a_0=-4.8\cdot 10^{-7}$,
  $a_1=594.1$, and $a_2=0.9994$.}
\label{Lanczos}
\end{figure}

The quantum critical point QCP1 at $\tilde{\alpha}=-4$ is characterized by a
level crossing of a singlet and a fully polarized state. Right at the
critical point the singlet-triplet gap $\Delta_{st}$ is therefore
expected to vanish. This is confirmed by the Lanczos calculations for
finite size chains shown in Fig.~\ref{Lanczos}.
We find, in particular, that the singlet-triplet gap vanishes as
$\Delta_{st}\sim N^{-5}$.

The thermodynamics of the $J_1-J_2$ model has been studied previously
by a Green's function method\cite{HaertelRichter} and by
TMRG.\cite{SirkerMF} In the second approach the one-dimensional
quantum model is mapped onto a two-dimensional classical model with
the help of a Trotter-Suzuki decomposition. It is then possible to
express the partition function in terms of a transfer matrix for the
classical model with the free energy depending only on the largest
eigenvalue of this transfer matrix.  The transfer matrix is extended
in imaginary time direction---corresponding to a successive lowering
of the temperature---with the help of a density-matrix renormalization
group algorithm. For details concerning the algorithm the reader is
referred to
Refs.~[\onlinecite{Peschel,BursillXiang,WangXiang,Shibata,SirkerKluemperEPL,SirkerKluemperPRB}].

\begin{figure}
\includegraphics*[width=.95\columnwidth]{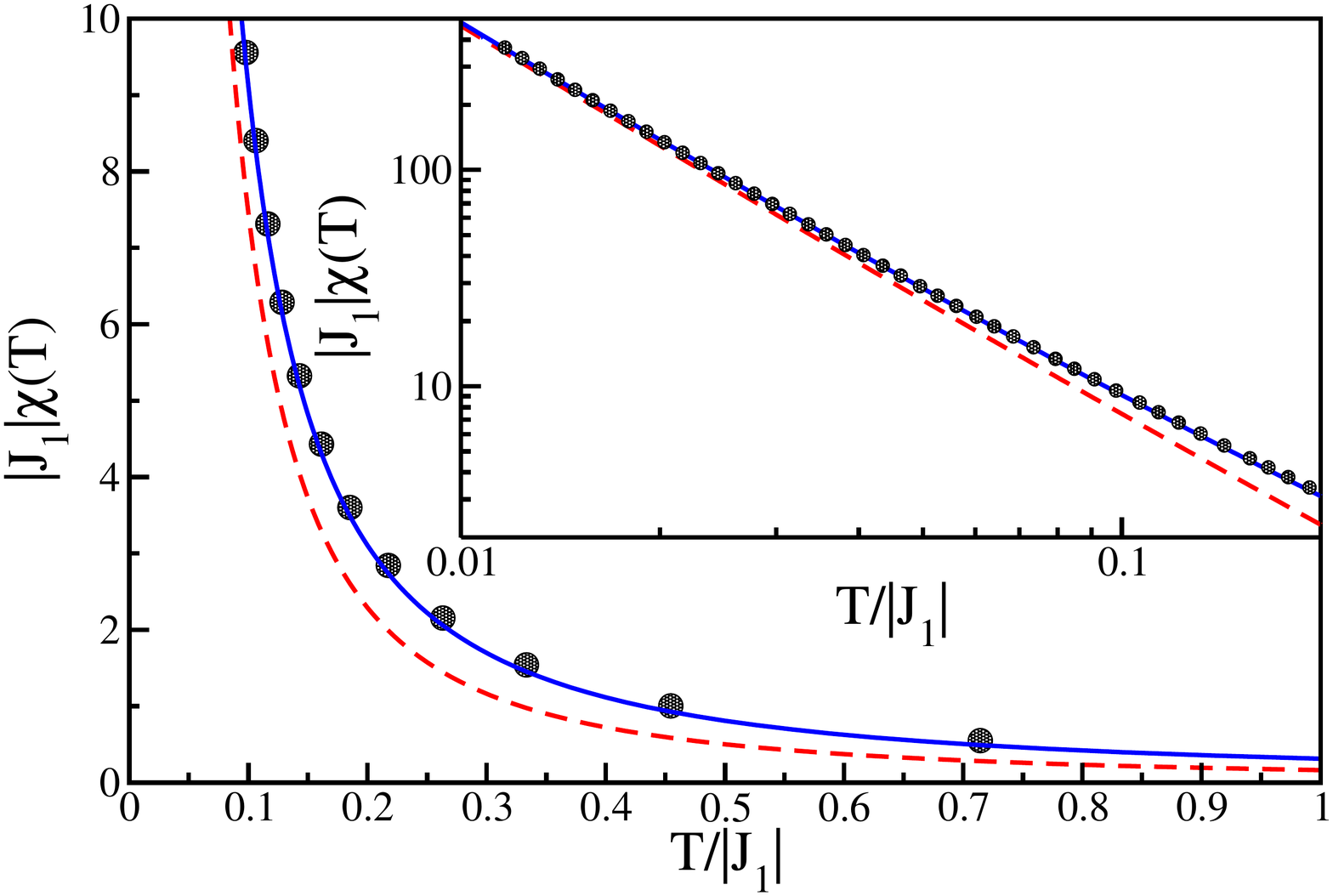}
\caption{(Color online) Susceptibility for $\tilde{\alpha}=-20$. The
  symbols denote the TMRG data, the solid line a solution of
  (\ref{MSWT.5}) where the Lagrange parameter $v$ is determined by
  solving the non-linear equations numerically, and the dashed line
  the leading low-temperature asymptotics (\ref{MSWT.6}) extracted
  analytically.  The inset shows the low-temperature region on a
  logarithmic scale.}
\label{Num.fig1}
\end{figure}

\begin{figure}
\includegraphics*[width=.97\columnwidth]{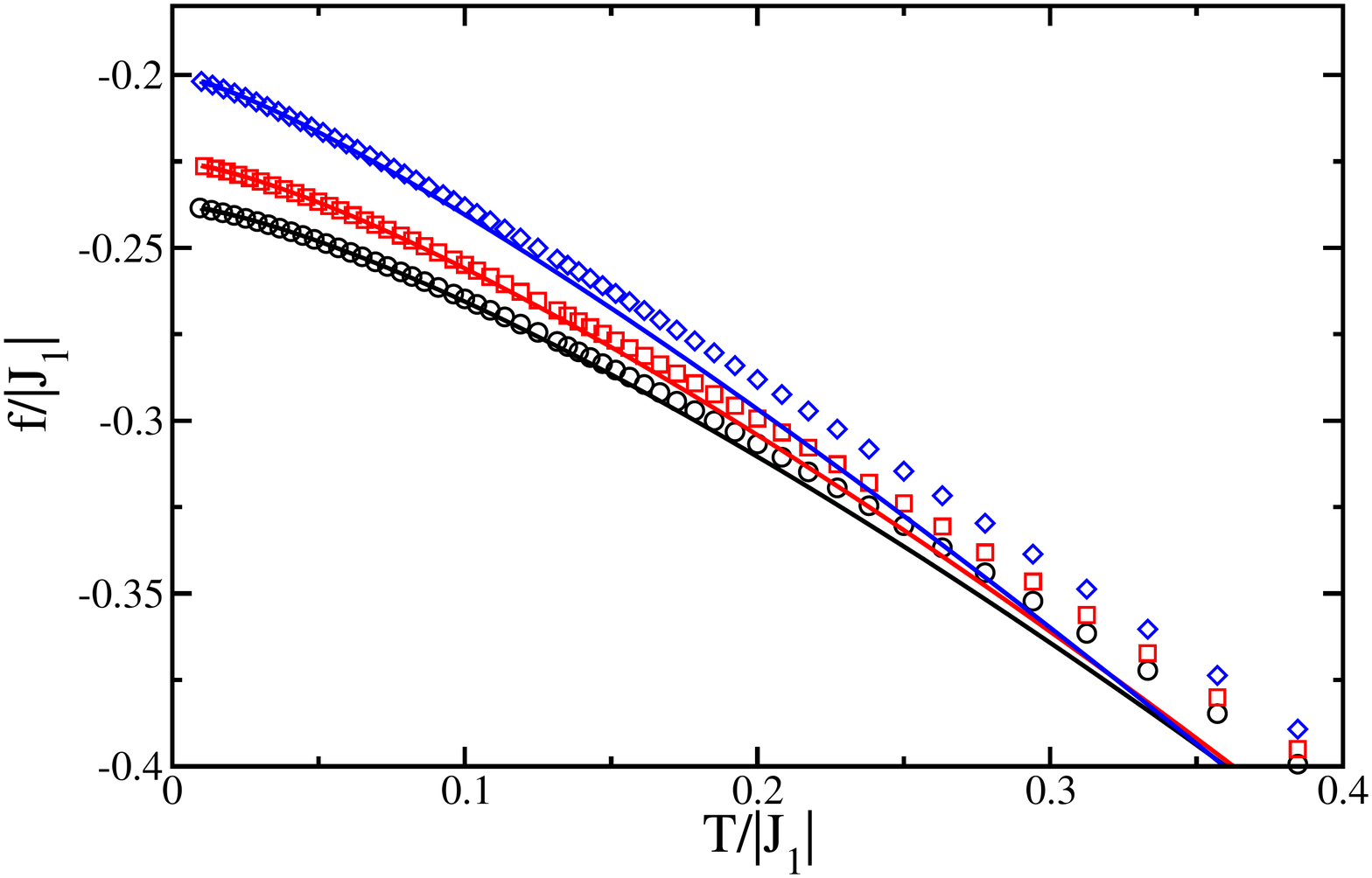}
\caption{(Color online) The free energy for $\tilde{\alpha} = -20,-10,-5$
  (from bottom to top). The symbols denote the TMRG data, the solid
  lines the MSWT result (\ref{MSWT.3}) using a fully self-consistent
  solution for the Lagrange parameter $v$.}
\label{Num.fig1b}
\end{figure}

\begin{figure}
\includegraphics*[width=.97\columnwidth]{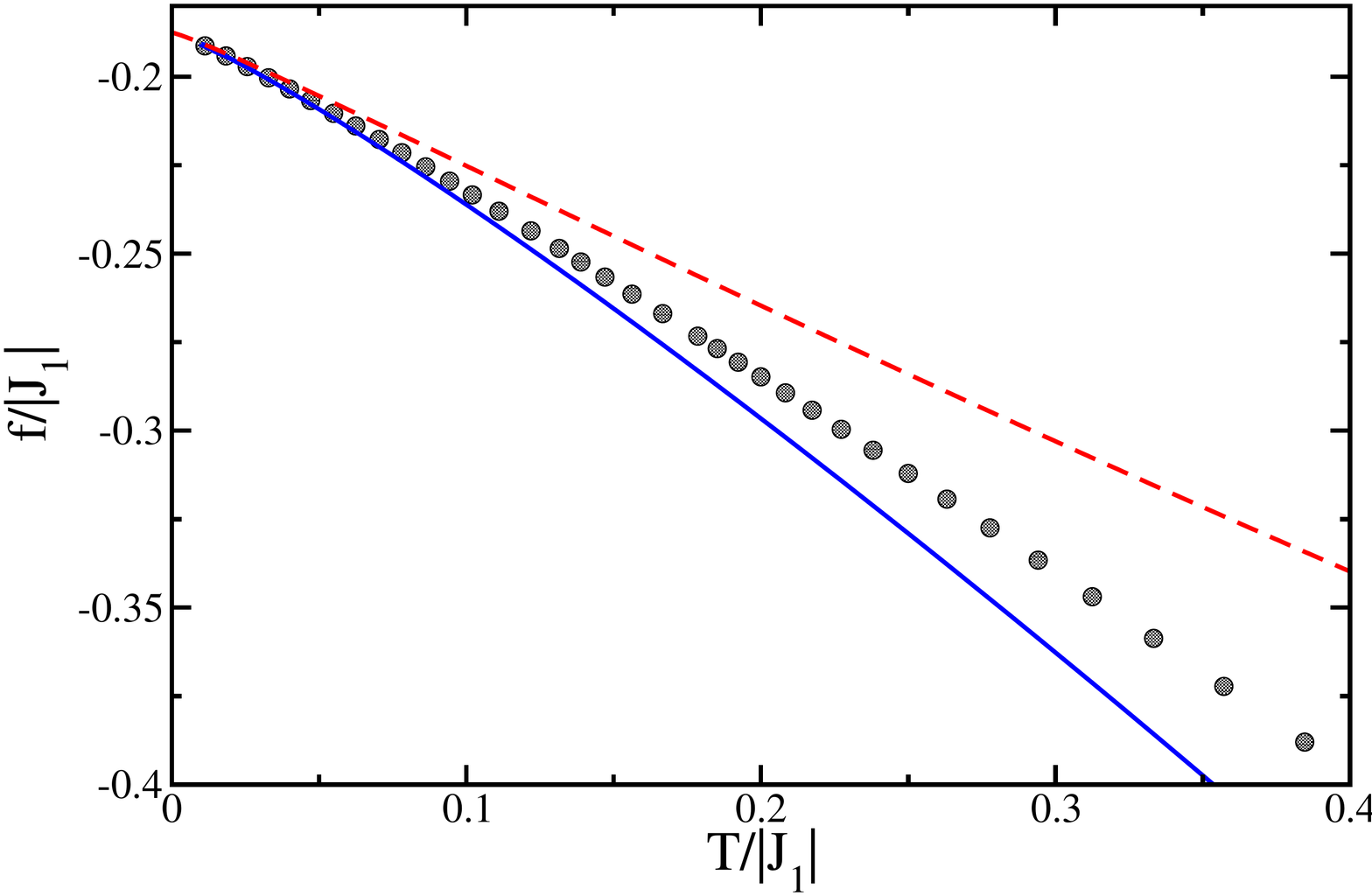}
\caption{(Color online) Free energy for the critical point
  $\tilde{\alpha}=-4$. The symbols denote the TMRG data, the dashed line the
  leading terms in the low-temperature asymptotics (\ref{MSWT.8})
  obtained by MSWT. The solid line represents the fully
  self-consistent solution of Eq.~(\ref{MSWT.3}).}
\label{Num.fig2a}
\end{figure}

Here we want to use the TMRG algorithm to test in how far the
analytical predictions from the previous section hold for the $s=1/2$
case. In Fig.~\ref{Num.fig1} numerical data for the susceptibility are
compared to MSWT for $\tilde{\alpha} =-20$.
If we numerically solve the non-linear equation for the Lagrange
parameter $v$ then the MSWT prediction is in excellent agreement with
the numerical data up to temperatures of order $t/s\sim 1$.
The formula (\ref{MSWT.5}) makes use of a representation of the
susceptibility in terms of the spin-spin correlation function
$\langle\vec{S}_i\vec{S}_j\rangle$ which also includes terms quartic
in the bosonic operators. 
In this case, the constraint fortunately makes it possible to obtain a
final expression which is still only bilinear in the bosonic
operators. 
The result presented in Fig.~\ref{Num.fig1} therefore goes
beyond linear spin-wave theory. The formula for the free energy,
Eq.~(\ref{MSWT.3}), is, however, a linear spin-wave expression. 
Here it is not possible to include the quartic terms without further
approximations because the constraint alone is not sufficient to
obtain a final expression which is only bilinear in the bosonic
operators.
The MSWT results for the free energy are therefore only valid at very
low temperatures as can be seen by the comparison with numerical data,
Fig.~\ref{Num.fig1b}.
A similar comparison for the critical point is shown in
Fig.~\ref{Num.fig2a}.
There is good quantitative agreement at temperatures $\tilde
t/s\lesssim1$, i.e. $T/|J_1|\lesssim 0.0625$ between the numerics and
the fully self-consistent solution of the MSWT equations.

\begin{figure}
\includegraphics*[width=.97\columnwidth]{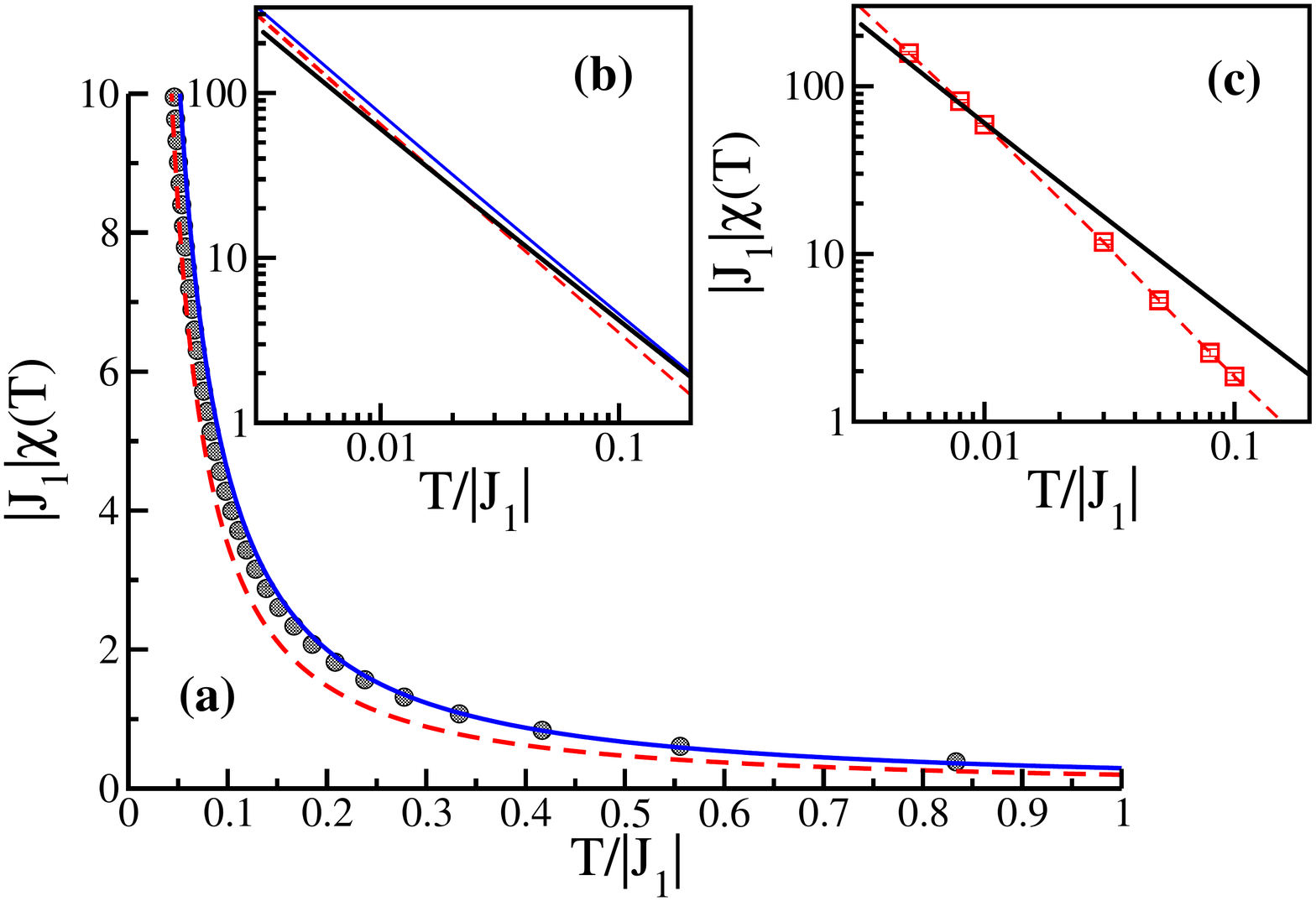}
\caption{(Color online) Susceptibility for the critical point
  $\tilde{\alpha}=-4$. (a) and (b): The symbols denote the TMRG data,
  the solid line a solution of (\ref{MSWT.5}) where the Lagrange
  parameter $v$ is determined self-consistently, and the dashed line
  the low-temperature asymptotics (\ref{MSWT.9}). The inset (b) shows
  the low-temperature region on a logarithmic scale. (c) Comparison
  between numerical data for the quantum model (small circles) and the
  classical model (squares).}
\label{Num.fig2}
\end{figure}

For the susceptibility the situation is expected to be more complex.
The scaling hypothesis used to derive Eq.~(\ref{susci_crit}) is
questionable because at the critical point we are above the upper
critical dimension. Indeed, we have already seen that the predictions
from MSWT deviate from formula (\ref{susci_crit}) which we have
confirmed to be the correct result for the classical model. This is
contrary to the ferromagnetic regime where the MSWT results coincide
at low temperatures with the solution of the classical model. A
comparison with numerical data, Fig.~\ref{Num.fig2}, indicates
nevertheless an apparently good quantitative agreement with MSWT up to
temperatures $T\sim |J_1|$.
A closer inspection of the low-temperature asymptotics (see inset (b)
of Fig.~\ref{Num.fig2}), however, shows that MSWT is not fully
consistent with the numerics. Furthermore, inset (c) shows that
classical and quantum model no longer share the same low-temperature
properties. If we fit the numerically obtained $\chi(T)$ for the
quantum model to a simple power law and vary the fit region we obtain
the values summarized in table \ref{Num.tab1}.
\begin{table}
\begin{ruledtabular}
\begin{tabular}{lcc}
$T_{\rm max}$ & $A$ & $\gamma$\\
\hline\\*[-0.15cm]
0.05 &0.245  & 1.195 \\
0.025 &0.259  & 1.185 \\
0.01 & 0.247  & 1.193 \\
0.0075 & 0.244  & 1.195 \\
0.005 &0.240  & 1.198\\
0.004 &0.237  & 1.201\\
\end{tabular}
\end{ruledtabular}
\caption{Parameters obtained by fitting $\chi(T)$ in Fig.~\ref{Num.fig2} to $\chi(T)=A\cdot T^{-\gamma}$ in the interval $T/|J_1|\in [0.003,T_{\rm max}]$.}
\label{Num.tab1}
\end{table}
This seems to indicate that the exponent might actually be smaller
than $4/3$ and therefore different from the exponent in the classical
model. However, using the TMRG algorithm we are not able to reach
temperatures for the quantum model which are as low as those
obtainable for the classical model using MC simulations. We therefore
can not completely rule out that the temperatures are just not low
enough to observe the $T^{-4/3}$ power law predicted by MSWT.

\section{Conclusions}
\label{Concl}
In summary, we have studied the frustrated $J_1$-$J_2$ Heisenberg
chain at and near to its $z=4$ critical point at $\tilde{\alpha}=J_1/J_2=-4$.
By developing a field theory we have discussed how the system is
driven from a ferromagnetic state to a state with incommensurate
(spiral) spin-spin correlations.  Based on this analysis, the
classical and the quantum model are expected to show the same
low-energy properties in the ferromagnetic phase.

From scaling arguments we obtained, in particular, that the
susceptibility diverges $\chi\sim (J_1+4J_2)/T^2$ in the ferromagnetic
phase with a known proportionality constant. We could reproduce this
result using an alternative analytical approach based on a modified
spin wave theory.  Furthermore, we have verified the analytical
prediction using Monte Carlo simulations for the classical model and
the transfer matrix renormalization group for the quantum model and
have found excellent agreement.

Right at the critical point, $J_1/J_2=-4$, simple dimensional analysis
allowed us to predict the scaling of the free energy and specific heat
with temperature. In particular, we found that the free energy scales
as $f\sim T^{5/4}$. Our modified spin wave theory calculations have
confirmed this scaling and the obtained parameter-free results have
been shown to be in good agreement with numerical data for the quantum
model at low temperatures.  Scaling arguments can also be used to
obtain a parameter-free formula for the low-temperature behavior of
the susceptibility, which---according to this formula---diverges as
$\chi\sim T^{-4/3}$. From the field theory analysis even the prefactor
can be obtained and we have shown, using Monte-Carlo simulations, that
this formula is indeed correct for the classical model.

The most interesting problem is the temperature dependence of the
susceptibility at the critical point $J_1/J_2=-4$ for the quantum
model. The modified spin wave approach also yields a $T^{-4/3}$
divergence, however, the prefactor is different from the one obtained
from field theory. The numerical data, furthermore, seem to indicate
that even the exponent might deviate from $4/3$. From fits of our
numerical data at the lowest accessible temperatures we have obtained
$\chi\sim T^{-1.2}$.

There are a number of possible reasons for this deviation: The
simplest explanation is that we do not have numerical data for low
enough temperatures to observe the true scaling behavior. However,
there are good reasons to believe that the observed deviation has
physical reasons. An inspection of the quartic term describing the
interaction of spin waves within spin wave theory shows that this term
is ultraviolet divergent. Such divergences are expected because
$d+z=5$ is larger than the upper critical dimension. In such a case
the ultraviolet properties of the theory can affect the critical
behavior.

Another problem is the treatment of the Berry phase term which we have
ignored in our field theory analysis. On the basis of a bosonization
approach,\cite{NersesyanGogolin,ItoiQin2} where the system is
considered starting from the decoupling point $\tilde{\alpha}=0$, the
spiral phase of the quantum model has been found to be gapped. Since
we do not expect an additional phase transition, a gap should exist
all the way to $\tilde{\alpha}\to -4$. In the field theory approach
this gap in the quantum model must be related to the Berry phase term.
This term might therefore also be important for the physical
properties right at the transition point. 
This expectation seems to be consistent both with the known degeneracy
of the ferromagnetic and resonance valence bond states at this point
and our numerical results which show that the low-temperature
properties of the classical and the quantum model are different.

The results attained here might be relevant and should be compared to
future experiments on edge-sharing cuprate chain compounds.

\begin{acknowledgments}
  J.S.~thanks I. Affleck for valuable discussions and acknowledges
  support by the MAINZ (MATCOR) school of excellence and the DFG via
  the SFB/Transregio 49. J.R., S.N.~and S.-L.D. thank the DFG for
  financial support (grants RI615/16-1 and DR269/3-1, respectively).
\end{acknowledgments}


\end{document}